\documentclass[12pt,preprint]{aastex}
\newcommand{\etal}{\mbox{\rm et al.}}

\newcommand{\msun}{\mbox{$M_{\odot}$}}
\newcommand{\rsun}{\mbox{$R_{\odot}$}}

\newcommand{\mearth}{\mbox{$M_\oplus$}}
\newcommand{\rearth}{\mbox{$R_\oplus$}}

\newcommand{\vsini}{\mbox{$v \sin i$}}

\newcommand{\teff}{\mbox{$T_{\rm eff}$}}
\newcommand{\fe}{\mbox{\rm [Fe/H]}}
\newcommand{\logg}{\mbox{${\rm \log} g$}}
\newcommand{\rhk}{\mbox{$\log R^\prime_{\rm HK}$}}
\newcommand{\shk}{\mbox{$S_{\rm HK}$}}

\newcommand{\mjup}{\mbox{$M_{\rm Jup}$}}
\newcommand{\rchisq}{\mbox{$\chi_{\nu}^2$}}     
\newcommand{\msini}{\mbox{$M\sin i$}}           
\newcommand{\ms}{\mbox{m s$^{-1}$}}

\newcommand{\kms}{\mbox{km s$^{-1}$}}
\newcommand{\rms}{\mbox{rms}}
\newcommand{\snr}{\mbox{\rm signal-to-noise ratio}}
\newcommand{\caii}{\ion{Ca}{2} H \& K}
  
\received{}
\accepted{}
  
\slugcomment{submitted to ApJ,  Sept 10, 2011}
  
\shortauthors{Fischer \etal}
\shorttitle{HIP~57274}
\begin{document}

\title{M2K: II. A Triple-Planet System Orbiting HIP~57274\altaffilmark{1}}
\author{Debra Fischer\altaffilmark{2},
Eric Gaidos\altaffilmark{3},
Andrew Howard\altaffilmark{4},
Matthew Giguere\altaffilmark{2},
John A. Johnson\altaffilmark{5},
Geoffrey W. Marcy\altaffilmark{4},
Jason T. Wright\altaffilmark{6,7},
Kelsey I. Clubb\altaffilmark{4},
Howard Isaacson\altaffilmark{4},
Kevin Apps\altaffilmark{8},
Sebastien Lepine\altaffilmark{9}, 
Andrew Mann\altaffilmark{10},
John Moriarty\altaffilmark{2},
John Brewer\altaffilmark{2},
Julien Spronck\altaffilmark{2},
Christian Schwab\altaffilmark{2},
Andrew Szymkowiak\altaffilmark{2}
}
  
\email{debra.fischer@yale.edu}

\altaffiltext{1}{Based on observations obtained at the Keck Observatory,
which is operated by the University of California}

\altaffiltext{2}{Department of Astronomy, 
Yale University, New Haven, CT 06511, USA}

\altaffiltext{3}{Department of Geology and Geophysics, University 
of Hawaii, Honolulu, HI 96822, USA}
  
 \altaffiltext{4}{Department of Astronomy, 
University of California, Berkeley, Berkeley, CA 94720, USA}

\altaffiltext{5}{Department of Astronomy, 
California Institute of Technology, Pasadena, CA 91125, USA} 

\altaffiltext{6}{Center for Exoplanets and Habitable Worlds, 525 Davey Lab, The Pennsylvania State University, 
University Park, PA 16803.}

\altaffiltext{7}{Department of Astronomy and Astrophysics, 525 Davey Lab, The Pennsylvania 
State University, University Park, PA 16803.}

\altaffiltext{8}{75B Cheyne Walk, Surrey, RH6, 7LR, United Kingdom}

\altaffiltext{9}{American Museum of Natural History
New York, NY 10023, USA}

\altaffiltext{10}{Institute for Astronomy, University of Hawaii, Honolulu, HI 96822, USA}

\begin{abstract}
Doppler observations from Keck Observatory have revealed a triple planet system orbiting the nearby mid-type K~dwarf, 
HIP~57274. The inner planet, HIP~57274b, is a super-Earth with \msini\ = 11.6 \mearth (0.036 \mjup), 
an orbital period of 8.135 $\pm$ 0.004~d, 
and slightly eccentric orbit $ e = 0.19 \pm 0.1$. We calculate a transit probability of 6.5\% for the inner planet. The 
second planet has \msini\ = 0.4 \mjup\ with an orbital period of 32.0 $\pm 0.02$~d in a nearly circular orbit, and 
$e = 0.05 \pm 0.03$. The third planet has \msini\ = 0.53 \mjup\ with an orbital period of 432 $\pm 8$~d (1.18 years) and 
an eccentricity $e = 0.23 \pm 0.03$.  This discovery adds to the number of super Earth mass planets 
with $\msini < 12 \mearth\ $ that have been detected with Doppler surveys.  We find that  
56 $\pm 18$\% super-Earths  are members of multi-planet systems. This is certainly a lower limit because of 
observational detectability limits, yet significantly higher than the 
fraction of Jupiter mass exoplanets, $20 \pm 8$\%, that are members of Doppler-detected, multi-planet systems.

\end{abstract}


\section{Introduction}
Low mass K and M~dwarf stars are important targets for exoplanet surveys because of their proximity and prevalence in the 
Galaxy.  Differences in the number and type of exoplanets orbiting these stars relative to more massive stars 
reflect conditions in the protoplanetary disk that are important for planet formation theory. Microlensing surveys suggest 
that both ice and gas giant planets are common at separations beyond the ice line \citep{g10}. However, the 
fraction of gas giant planets detected inside the ice line by Doppler surveys is relatively low for late K and 
early M dwarfs \citep{e03, b06}. \citet{fg11} find that the population of giant planets has a 
precipitous decline for stars redward of \bv = 1.1, a spectral type of about K5V. \citet{c08} estimate that relative to 
FGK stars, M~dwarfs are far less likely to harbor gas giant planets with periods shorter than 5 years. 
\citet{j10} find that $3.4^{+2.2}_{-0.9}$\% of low mass stars ($M < 0.6 M_\odot$) have planets with 
$\msini > 0.3 \mjup$ and semi-major axes less than 2.5 AU. 

The remarkable discovery of more than 1200 planet candidates by the Kepler Mission \citep{bor11} provides 
statistics for smaller planets, and suggests that the reduced planet occurrence with later spectral type
only applies to gas giant planets. After correcting for the poorer detectability of transits around 
higher mass stars with larger radii, \citet{h11} find that 20 - 30\% of low mass stars have planet 
candidates with Neptune-like radii between 2 - 4 \rearth\ while the fraction of more easily detected 
Jupiter-radii planets hovers at a few percent.  Howard et al. also see evidence of a rising occurrence of 
small-radius planets among cooler, less massive stars. Further, \citet{sl11} find that while 
the planet-metallicity correlation among sun-like stars is strongest for those hosting large-radius planets, the 
planet-metallicity correlation among low-mass stars is significant even among hosts of small-radius planets. 
Doppler surveys of nearby, low-mass stars provide a means of testing whether these correlations hold among 
stars in the solar neighborhood, and if so, inform the target lists of future planet search efforts.  

Late type stars are especially appealing targets for rocky planet searches in Doppler surveys because the lower stellar mass 
results in a larger reflex velocity for a given mass planet. Furthermore, chromospheric activity in low mass stars has less 
impact on the radial velocities \citep{if10, l11}.  The ubiquity of low mass stars, coupled with more easily detected Doppler 
signals and lower stellar jitter all make K and early M dwarfs desirable targets in the search for rocky planets. However, 
a caveat has emerged: the inner planetary architectures of low mass stars may be more complex. \citet{lath11} analyzed 
multi-planet systems detected in transit with the Kepler Mission and find that solar type and hotter stars are more 
common hosts of single transiting planets, while multi-planet systems are more 
often detected around cooler stars.  
Among the 170 multi-planet systems detected by the Kepler Mission, 78\% contain planets no larger 
than Neptune; close-in gas giant planets are far less common in multi-planet systems \citep{lath11}.  This has profound implications for Doppler surveys: the challenge of detecting the small velocity amplitudes of Neptune-like planets 
will be compounded by the need to deconvolve multiple signals of similar amplitude. For both of these 
reasons, a larger number of observations over a longer interval of time
are required to resolve the components of these planetary systems. 

To better understand the frequency and architectures of planetary systems around low mass stars, we 
began ``M2K'' \citep{a10}, a Doppler survey of M and K dwarf 
stars drawn from the SUPERBLINK proper motion survey \citep{ls05, lg11}. Here, we report the detection 
of a triple planet system orbiting HIP~57274 comprised of a super Earth-mass planet and two planets that 
are likely gas giants. 

\section{Observations and Data}
Doppler observations are carried out with the Keck 10-m telescope and 
the HIRES spectrograph \citep{v94}.  An iodine cell is used to provide the wavelength solution and sampling of the 
line spread function to model the Doppler shift in the stellar spectra \citep{b96}. The B5 decker on HIRES provides a 
spectral resolution of about 55,000 and an exposure meter terminates the observations when a target \snr\ of about 
200 is achieved. Most of the M2K stars are fainter than $V=9$, requiring exposure 
times of 10 - 15 minutes.  We have acquired three or more 
Doppler measurements for more than 170 stars, with formal measurement uncertainties of about 1.5 \ms. 

\subsection{HIP57274} 
HIP 57274 (GJ~439) has an apparent magnitude of $V=8.96$, color \bv\ = 1.111, and parallax of 38.58 $\pm 1$ mas 
according to the {\it Hipparcos} catalog \citep{esa97, vanleeuwen07}.  This
yields a distance of 25.9 pc and absolute visual magnitude 
of $M_V = 6.89$. We carried out spectral synthesis modeling of the iodine-free template spectrum 
using Spectroscopy Made Easy \citep{vp96, vf05} to determine stellar parameters.  Following the method 
described in \citet{v09}, the initial parameters derived with SME were used as input for interpolation of the 
Yonsei-Yale (Y2) isochones \citep{d04}, which returns a new value for \logg. We then ran an iterative loop, fixing \logg\ to the isochrone value and running a new SME model fit.  The other (free) stellar parameters change in response to the fixed surface gravity, so subsequent isochrone interpolations produce a slightly different value for \logg.  We continue the iteration until the output \logg\ form the isochrones does not change by more than 0.001 dex from the previous iteration. 
This provided the following spectroscopic parameters: 
\teff\ = 4640 $\pm 100$K, \vsini\ = 0.5 $\pm 0.5$ \kms, \fe\ = +0.09 $\pm 0.05$, \logg\ = 4.71 $\pm 0.1$, and Y2 
isochrone models for a stellar mass of 0.73 $\pm 0.05$ \msun, radius of 0.68 $\pm 0.03$ \rsun\ and an age of 
7.87 $\pm 5$ Gyr.  The spectral classification is listed as K8V in the {\it Hipparcos} catalog, although the \bv\ color, 
spectroscopic temperature and derived mass are more consistent with a slightly
earlier spectral type. \citet{g03} list a spectral classification of K4V for this star.  A medium-resolution 
spectrum, obtained by one of us (SL) with the Mark III spectrograph at the MDM 1.3-m telescope, returns 
a spectral type of K5V, and we adopt this spectral classification for this paper.  The stellar parameters for 
HIP~57274 are summarized in Table 1. 

\subsection{Chromospheric Activity and Velocity Jitter}
\citet{if10} determined the chromospheric activity of 2630 stars observed at Keck by measuring the 
emission in the cores of the \caii\ lines relative to adjacent continuum regions.  
These \shk\ values were calibrated to the long-standing \shk\ values from the 
Mt. Wilson program \citep{d91}. Using their technique, we measure a mean \shk\ = 0.38 for HIP~57274. The individual measurements of \shk\ are listed in the last column of Table 2, 
along with the radial velocity measurements. 

Cooler stars typically have larger \shk\ values than solar type stars because 
of weaker continuum in the near-UV.  Therefore, \shk\ values should not be directly compared for stars of different spectral 
types.  \citet{n84} correct for the photospheric contributions to produce a normalized activity metric, \rhk.
Chromospheric activity is tied to dynamo-driven flows and decreases as the star ages and spins down. 
\citet{n84} made use of this activity-rotation correlation and calibrated \rhk\ to rotational periods for 
stars in open clusters of different ages.  Using their relation, we derive  \rhk\ = -4.89, indicating low activity for 
HIP~57274 and ${\rm P_{rot} \sim 45}$~d, consistent with the SME-estimated age of $\sim 8$ Gyr. 
We caution that both \rhk\ and ${\rm P_{rot}}$ were 
calibrated by \citet{n84} using stars with  $0.4 <  \bv\ < 1.0$ and rotational periods shorter than 30 days; HIP~57274 falls outside 
both of these properly calibrated ranges and therefore our derived \rhk\ and rotational period should be considered 
to be less certain estimates. 

Because the activity calibration for stars redward of \bv\ = 1.0 is an extrapolation, \citet{if10} established a 
differential activity measurement, $\Delta S_{HK}$, and evaluated the impact of chromospheric activity on 
radial velocities in four separate ranges of \bv. 
Following their method, we plot the \shk\ index for the 170 stars observed on the M2K program with \bv\ color from 
0.8 to 1.6 (Figure 1).  The dashed line in this plot is taken from taken from \citet{if10} and indicates the 
baseline \shk\ values for low activity stars.  $\Delta S_{HK}$ is the difference between this baseline activity level 
and the mean \shk\ for a given star.  Active stars float high above the baseline values while chromospherically quiet 
stars are closer to the dashed line in Figure 1. 

In Figure 2, we plot velocity \rms\ as a function of excess chromospheric activity, $\Delta S_{HK}$ for all of the 
observed M2K stars. We fit a 
linear function to the lower twentieth percentile velocity scatter and interpret this (red solid line overplotted in Figure 2) 
as the quadrature sum of internal errors and jitter (where jitter is both instrumental and astrophysical).
The \rms\ scatter for inactive stars with $\Delta S_{HK} \sim 0.0$ is 2.38 \ms, and given the typical internal errors of 1.5 \ms, 
this implies a minimum jitter of 1.45 \ms. We measure $\Delta S_{HK}$ = 0.03 for HIP~57274, suggesting a low 
stellar jitter of $\sim1.5$ \ms.

\subsection{Doppler Observations and Keplerian Model}
Figure 3 shows the time series data, overplotted with our Keplerian model for a triple planet system.  We used the 
partially linearized Levenberg-Marquardt algorithm \citep{wh09} built into the Keplerian Fitting Made Easy (KFME) 
program \citep{mg11} to model the data. The best fit Keplerian model for three planets includes 
a trend of -0.026 \ms\ per day. Parameter uncertainties were calculated with a bootstrap Monte Carlo 
analysis  \citep{m05} in KFME  \citep{mg11}, which iteratively fits the data with a best-fit model, then adds the scrambled 
residuals back to the theoretical velocities before refitting.  

The planet wiht the shortest period completes one orbit in $8.135 \pm 0.004$ days and induces a velocity amplitude 
of $4.64 \pm 0.46$ \ms. With a stellar mass of 0.73 \msun, we derive a planet mass \msini\ = 11.6 \mearth and semi-major 
axis of 0.07 AU.  The orbital 
eccentricity is $0.187 \pm 0.10$ and the argument of periastron passage $\omega \sim 82^\circ$.  Because the 
velocity amplitude is small compared to the uncertainties and stellar jitter, the eccentricity for this planet is poorly 
constrained, however the signal itself is unambiguous. Figure 4 shows the periodogram of the residual velocities of 
HIP~57274 after removing the other two planets and linear trend described below.  We carried out a 
Monte Carlo test to determine the false alarm probability (FAP) of the periodogram power. 
In this test, 10,000 trials were carried out where  
the best fit (triple Keplerian and trend) model was subtracted and the residual velocities were scrambled before being 
added back to the theoretical velocities and refitting.  In these 10,000 trials, a peak of the same height was never found 
in the residuals to the fit of the two more massive planets, yielding a FAP less than $10^{-4}$. 
Figure 5 shows the phase-folded velocities of HIP~57274b overplotted with the theoretical Keplerian model 
after removing the signals from the two additional planets and the linear trend. 

We calculated the prospective time of transit, transit duration and transit probability using 
KFME \citep{mg11}.  The transit ephemeris is 2455801.779 $\pm 0.27$~HJD, or 2011/08/28 06:41:40.7~UT.
The next transit observable from Mauna Kea occurs at 3 AM HST on 13 January 2012, except that the 
uncertainty in the transit time is more than 6 hours. 
The duration of the prospective transit would be 3.08 $\pm0.356$ hours and the geometric probability that this 
planet will transit is 6.5\%. 

The middle planet in this system has an orbital period of $32.0 \pm 0.02$ days.  The best fit model for this planet has 
a nearly circular orbit, with eccentricity of $0.05 \pm 0.02$. The velocity amplitude is $32.4 \pm 0.6$ \ms, and 
implies a planet with \msini\ = 130 \mearth\ or 0.41 \mjup\ and orbital radius of 0.18 AU. The phase-folded data 
and model for HIP~57274c is shown in Figure 6 after subtracting theoretical velocities for the linear trend and 
the inner and outer planets. The prospective ephemeris time is 2454793.035 $\pm 0.176$ HJD, although 
the longer orbital period for the middle planet results in a lower 2.7\% transit probability.

The third planet, HIP~57274d, has an orbital period of $432 \pm 8$ days, a velocity amplitude of $18.2 \pm 0.8$ \ms, 
and orbital eccentricity of $0.27 \pm 0.03$.  The planet mass is \msini\ =0.53 \mjup, and the semi-major axis of the orbit 
is a familiar 1.01 AU.  Figure 7 shows the time series velocities and Keplerian model for this planet after removing the 
inner two planets and the linear trend. The velocity \rms\ to the triple-planet fit is 3.15 \ms.  However, if we adopt the 
predicted jitter of 1.5 \ms, we find that \rchisq\ = 2.7, indicating that the model does not fully describe our observations. 

To check for additional planets in the system, we subtracted the theoretical velocities for the linear trend and 
three Keplerian models.  Figure 8 shows a periodogram of the residual velocities, with significant peaks at 
1.019 d and 52.996~d that could be aliases of each other, due to the diurnal cadence: 1.0 + 1./52.996 = 1.019 d. 
This predicts the presence of a second peak at 1.0 - 1./52.996 - 0.981, and we see a second peak in the periodogram 
at 0.981 d. Interestingly, the $\sim 53$ day peak has been increasing in 
strength. Unlike the periodograms for the three planets that we modeled (HIP~57472b, c, and d), 
the 53-d peak is flanked by two additional peaks. 
We suspect that this signal may be caused by spots rotating on the surface of the star which reduce flux on the approaching 
blue-shifted edge of the star and then on the receding red-shifted edge of the star.  Physically, this would produce a 
line profile asymmetry that could be spuriously modeled as a Doppler shift.  We checked to see if the residual velocities to 
the full triple planet fit were correlated with the activity measurements but found only an insignificant trend Figure 9 
(dashed line). We tried detrending the velocities with the linear fit shown in Figure 9, however this only slightly reduced 
the periodogram power in the residual velocities. If we blindly fit this periodic signal with a 
Keplerian model, we derive a period of 53.2 days with an amplitude of 2.6 \ms, and the residuals drop 
to 2.63 \ms with a \rchisq\ = 1.28. Interpreting this signal as an additional noise source from coherent spots, we can 
add the 2.63 \ms\ signal in quadrature with the expected jitter of 1.5 \ms\ to obtain a revised jitter estimate of 2.8 \ms. This 
changes \rchisq\ to 1.06 for the model of a triple planet system plus a linear trend. 
If the 53 d signal originates from star spots, then it should have a detectable photometric signal. We have started a photometric campaign to check for spot modulation and to search for the transit signal of the inner 8.135 d planet.  In addition to the photometric observations, we will continue to obtain Doppler measurements to better understand the origin of the 53 day signal. 

 We have started a photometric 
campaign to check for spot modulation and to search for the transit signal of the short-period planet.  If the 53 d signal 
originates from star spots, it should have a detectable photometric signal. In addition to the photometry, we will continue
to obtain Doppler measurements to better understand the origin of the 53 day signal. 
 
\subsection{Dynamical stability}

To assess the stability of the HIP~57274 system, we ran an ensemble of dynamical simulations of the triple planet system 
with the hybrid symplectic integrator code Mercury 6 \citep{c99}.  Orbital parameters for each body were calculated 
from the values in Table 1 assuming Gaussian-distributed errors (with a truncation at 0.0 for eccentricity).  We ran 20 simulations 
with $\sin i = 1$ (minimum mass case) and 15 simulations with $\sin i = 0.3$ (95\% of random orientations will have 
a $\sin i$ greater than this value) and a mass of 0.73~M$_\odot$ for the central star.  For radius and thus collision 
probabilities, we assumed mean density of 6.0, 1.0, and 1.0 g~cm$^{-3}$ for the 
b, c, and d planets, which would be typical for the densities of rocky and gas giant planets.  The time step was set 
to 0.365~d (4.5\% of the orbital period of HIP~57274b) and each simulation was run for 10~Myr.  In none of the 35 independent
simulations did collisions or ejections occur.  We conclude that the system is stable regardless of its inclination.

\section{Summary and Discussion}

Here we present a triple planet system orbiting the late K~dwarf star, HIP~57274.  The inner planet orbits in 
8.135 days and has a mass, \msini\ = 11.6 \mearth.  The orbit is slightly eccentric with periastron directed toward
our line of site.  We calculate a transit probability of 6.5\% with a putative ephemeris 
times, $T_c =2455801.776 \pm 0.338$ and a duration of $3.08 \pm 0.35$ hours. 
The second planet orbits in 32 days and has \msini\ = 130.0 \mearth.  The third planet has an orbital period of 432 days with 
a semi-major axis of 1.01 AU and \msini\ = 167.4 \mearth. The nominal habitable zone of this K star, 
corresponding to 0.95-1.3 AU around the Sun (Kasting et al. 1993) and adjusting for the lower luminosity, 
lies between 0.41-0.57 AU, or orbital periods of 110-180d.   No significant periodic signals lie within this 
range (Figure 8).

With the addition of HIP~57274b, there are now 25 planets with \msini\ $< 12.0$ \mearth\ listed in the Exoplanet Orbit Database, or EOD \citep{w11}. It is probable that these 25 planets are super-Earths or Neptunes, rather than gas
giant planets.  Importantly, 
only 8 of the 25 (currently) appear to reside in single planet systems. The remaining 17 low mass planets are constituents 
of 10 multi-planet systems. Counting systems instead of planets and applying Poisson error bars (i.e., the percentage of 
single or multiple planet 
systems divided by the square root of the number of planets), we find that 44 $\pm 16$\% of these low mass planets are 
single (at the current level of Doppler detectability), while 55.6 $\pm 17.6$\% are in multi-planet systems.  To restate: more 
than half of the Doppler detected super-Earths  are detected as members of multi-planet systems.  

To compare the architectures of planetary systems containing super-Earths  with those containing gas giant planets, we 
extracted all 36 planets from Doppler surveys with \msini\  between 1.0 - 1.5 \mjup\ orbiting main sequence stars 
in the EOD. In this Jovian-mass subsample, 
there were 30 single planet systems (86 $\pm 16$\%) and 5 multi-planet systems (14~$\pm 6$\%). This result is not particularly 
sensitive to the arbitrary range of exoplanet mass: in a subsample of 63 planets with \msini\ from 1.5 - 2.5 \mjup, 73~$\pm 11$\% systems were single. Another sample cut of 41 planets with \msini\ from 3.0 - 6.0 \mjup\ yielded 83~$\pm 14$\% single 
planet systems. Taking an average of these three subsamples, roughly 80~$\pm 10$\% of the Doppler-detected Jupiters 
are currently in single planet systems, 
and about 20~$\pm 8$\% of Jupiters are in multi-planet systems. This can be compared with the estimate of \citet{w09} who find 
that at least 28\% of planets are in multiple systems.  Since they count planets of all masses detected before 2009, that result is not 
inconsistent with our estimate. Relative to super-Earths , Doppler surveys detect significantly fewer multi-planet systems 
with gas giant planets. 

In Figure 10, we consider the sibling planets in these multi-planet architectures and compare the super Earth and Jupiter 
subsamples. A total of 26 planets accompany the 10 super-Earths  in known multi-planet systems; these sibling planets 
also tend to be systematically low in mass - the median \msini\ = 23.5 \mearth.  In contrast, the Jovian subsample includes 
six companions, spanning a range from 0.58 \mjup\ to about 4 \mjup\ with a median \msini\ of 2.6 \mjup.  

Is the dramatic difference in the architecture of low mass planets a bias in Doppler detection efficiency or the result 
of nature (eg., conditions in the protoplanetary disk or evolutionary processes)?
In Figure 11, we compare the stellar hosts for the super Earth and Jovian subsamples. The histogram of stellar mass for the 
hosts of super-Earths  is offset to lower mass, with an average of 0.7 \msun, while the host stars of the Jovian sample have 
a mean stellar mass of 1.06 \msun.  The dependence of reflex velocity on stellar mass implies that the Doppler signals for 
a star with a mass of 0.7 \msun\ would be amplified by about 30\% relative to a star of 1.06 \msun. Therefore, the paucity of Jupiter like planets around lower mass stars \citep{fg11} cannot be a selection effect.  

However, assessing the scarcity of super-Earths  around the more massive host stars of Jupiter mass planets 
is complicated by 
observational detectability issues. The mean velocity amplitude of the super Earth sample would drop from 4.2 \ms\ for the 
mean host star mass of 0.7 \msun\ to 3.2 \ms\ around solar mass stars.  At the same time, as the stellar mass increases 
from 0.7 \msun\ to 1.06 \msun, the minimum stellar jitter increases by a factor of two \citep{if10,l11} or more for chromospherically active stars.  As a result, if stars with close-in Jovian mass planets also host a system 
of super-Earths, the Doppler signal would be roughly a $1 \sigma$ detection. 
Furthermore, the observational biases that influence second-planet detection are complex.  In some cases the presence of one planet can complicate the detection of additional, lower amplitude planets (for instance if the planets are in resonance \citep{alc10} if one planet has poorly constrained orbital parameters, or if the observational cadence causes aliased signals near the orbital frequency of the additional planet).  On the other hand, the presence of a planet can also cause a star to receive additional observations in preparation for publication, making the detection of a low-amplitude planets more likely.

\citet{r08} find that although the migration of giant planes does not completely impede terrestrial planet growth, 
the final accretion phase of terrestrial planets is affected by gravitational perturbations from gas giant planets. 
Although the Doppler detections may only weakly constrain the presence of low mass planet siblings to 
Jupiter mass planets, the Kepler data provide additional support for this case.  
\citet{lath11} proposed that the transiting Jovian planets detected by Kepler would have migrated into their current 
locations and likely destabilize the orbits of smaller planets.  Likewise, nearly all of the Jovian planets detected by Doppler 
surveys are migrated planets that may have driven out close-in Neptunes.

\acknowledgements
We gratefully acknowledge the dedication and support of the Keck Observatory staff, especially 
Grant Hill and Scott Dahm for support of HIRES and Greg Wirth for support of remote observing. 
DAF acknowledges research support from NSF grant AST-1036283 and NASA grant NNX08AF42G.  
This research has made use of the Exoplanet Orbit Database at exoplanets.org 
and NASAÕs ADS Bibliographic Services. Data presented herein were obtained at the W. M. Keck 
Observatory from telescope time allocated to Yale University and to the National Aeronautics
and Space Administration through the agency's scientific 
partnership with the California Institute of Technology and the 
University of California and to the University of Hawaii.  
This research has made use of the SIMBAD database, operated
at CDS, Strasbourg, France, and of NASA's Astrophysics Data System
Bibliographic Services.The Observatory was made possible by 
the generous financial support of the W. M. Keck Foundation.
The authors wish to recognize and acknowledge the very significant 
cultural role and reverence that the summit of Mauna Kea has always 
had within the indigenous Hawaiian community. We are most fortunate
to have the opportunity to conduct observations from this mountain.

\clearpage

\clearpage  

\begin{figure}
\plotone{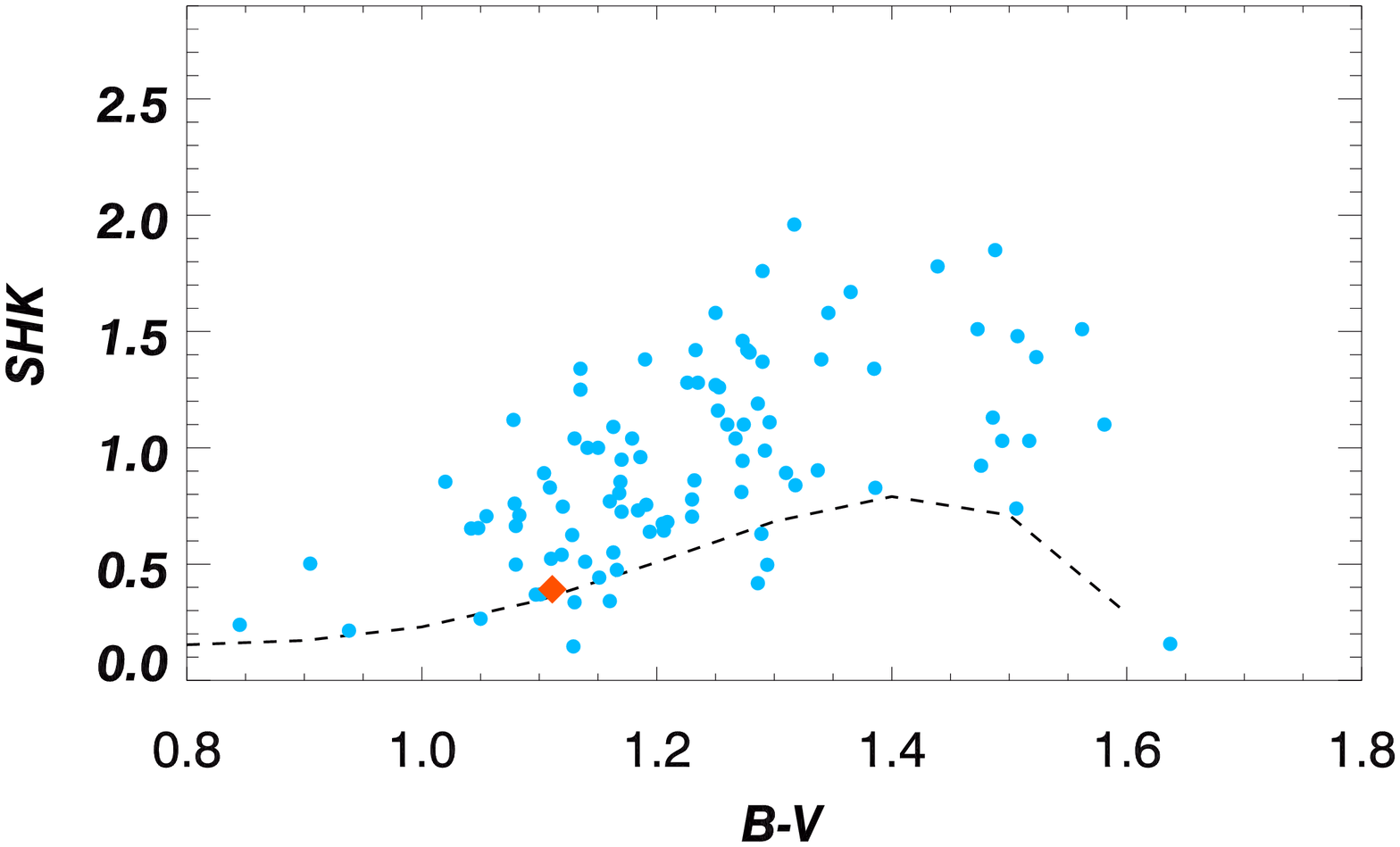}
\figcaption{The emission in the core of the \caii\ lines is a signature of chromospheric activity that is parameterized 
as a \shk\ value for each star and plotted above as a function of B-V. The dashed line represents the baseline for low 
activity stars from \citet{if10}.  Inactive stars fall near the dashed line with $\Delta S_{HK} \sim 0.0$ and 
active stars have a large $\Delta \shk$. A filled diamond is used to show the \shk\ measurement for HIP~57274. }
\end{figure}

\begin{figure}
\plotone{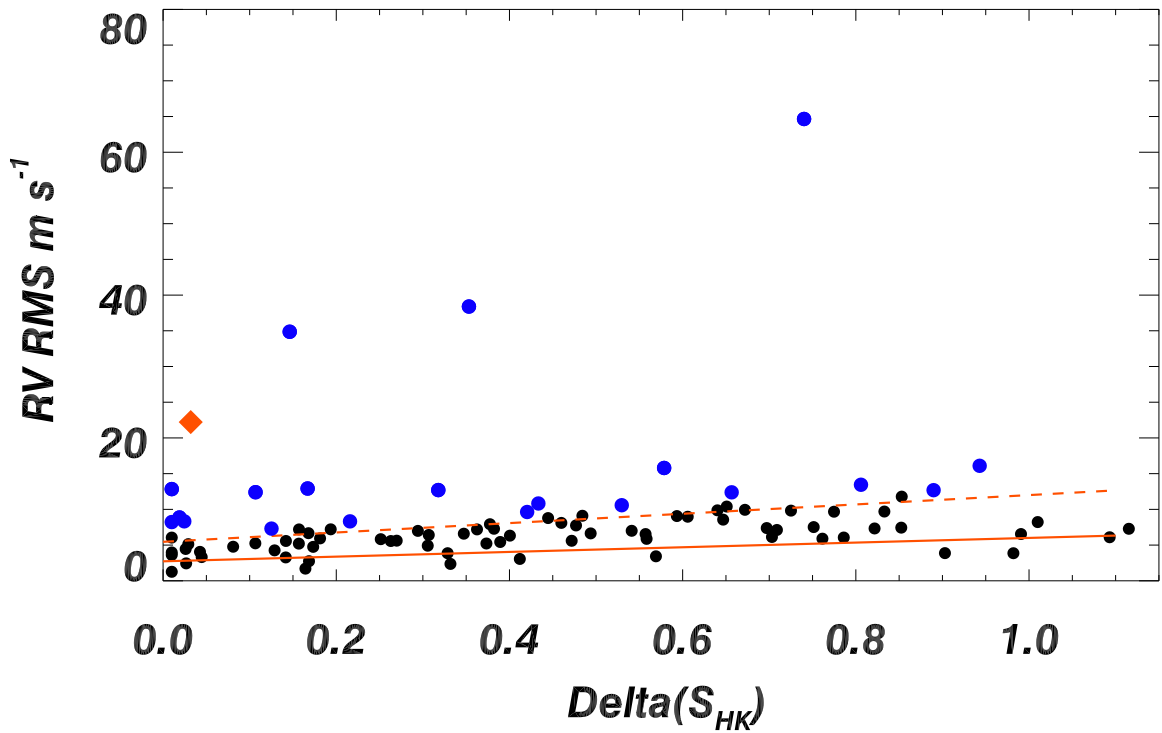}
\figcaption{The \rms\ velocity scatter is plotted as a function of activity. No trends in the velocity data were 
removed. The solid red line is a linear fit to the bottom twentieth percentile velocity scatter. Stars with more than 
$3 \sigma$ jitter (the dashed line) are indicated with green dots and represent prospective planet candidates. 
HIP~57274 is represented by the filled diamond.  }
\end{figure}

\begin{figure}
\plotone{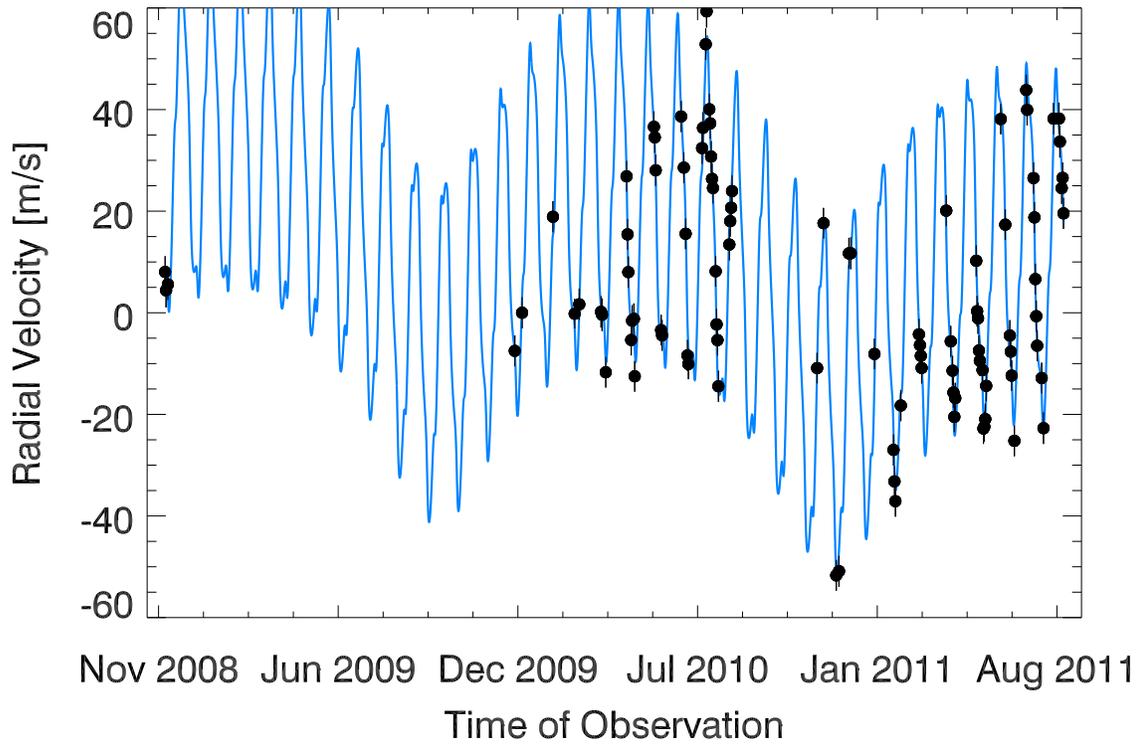}
\figcaption{The time series data for HIP~57274 are shown with a Keplerian model for 
three planetary signals plus a linear trend. The model fit has an \rms\ in the residual velocities 
of 3.15 \ms\ and \rchisq = 1.06, with an assumed jitter of 2.8 \ms\ added in quadrature to the internal errors.}
\end{figure}

\begin{figure} 
\plotone{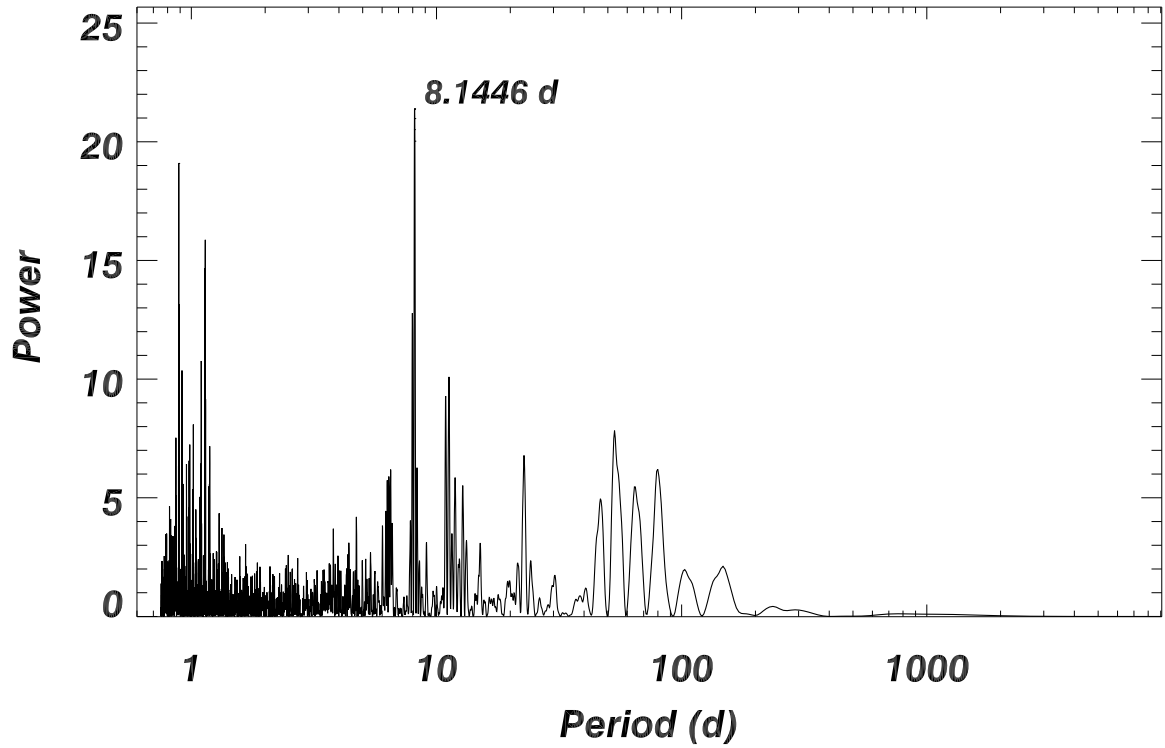}
\figcaption{After subtracting the linear trend and the best fit theoretical velocities for the two outer planets, the 
periodogram of the residuals shows signifiant power at 8.135 days. The signal of this inner planet, which we 
designate as HIP~57274b, has a FAP $< 0.0001$. The two peaks just below and above a period of 1.0 day 
are aliases of the 8.14 d peak}
\end{figure}

\begin{figure}
\plotone{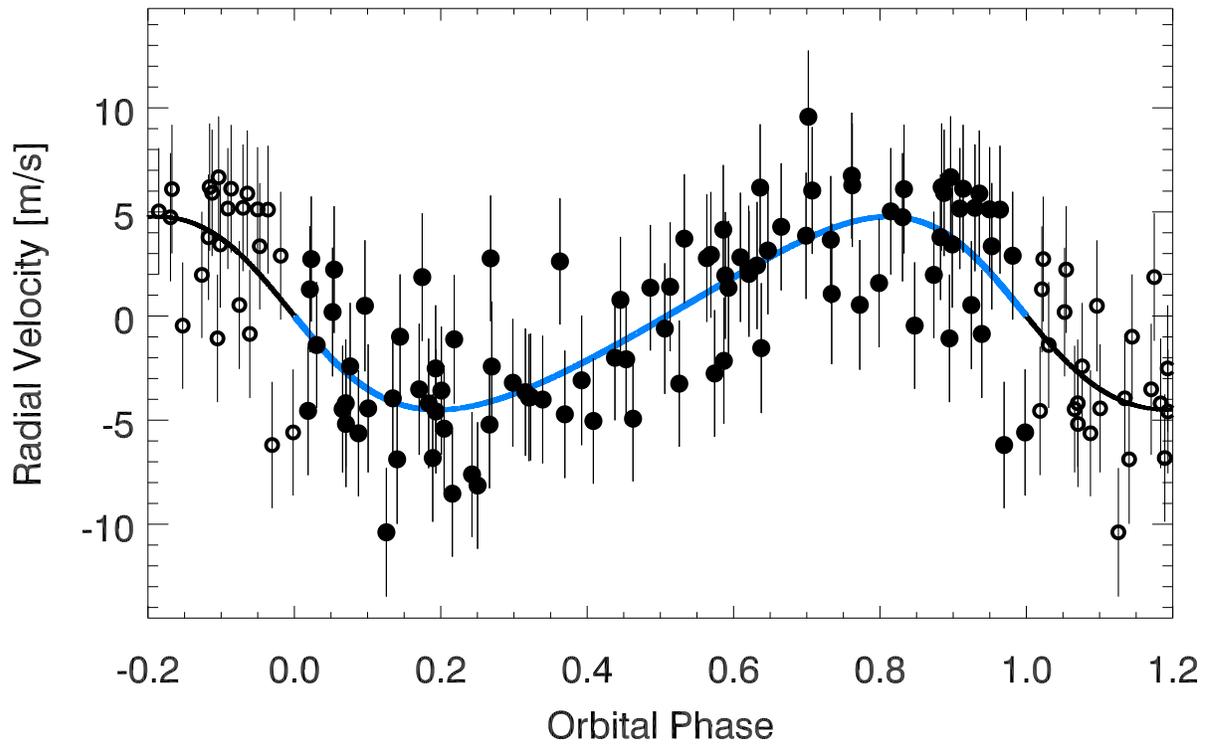}
\figcaption{The phase-folded radial velocities for HIP~57274b are shown after removing Keplerian signals from 
the outer two planets and subtracting a linear trend. The Keplerian model is plotted with a solid line and 
has an orbital period of $8.135 \pm 0.005$~d, 
orbital eccentricity $e = 0.19 \pm 0.1$ and \msini = 11.6 \mearth.}
\end{figure}

\begin{figure} 
\plotone{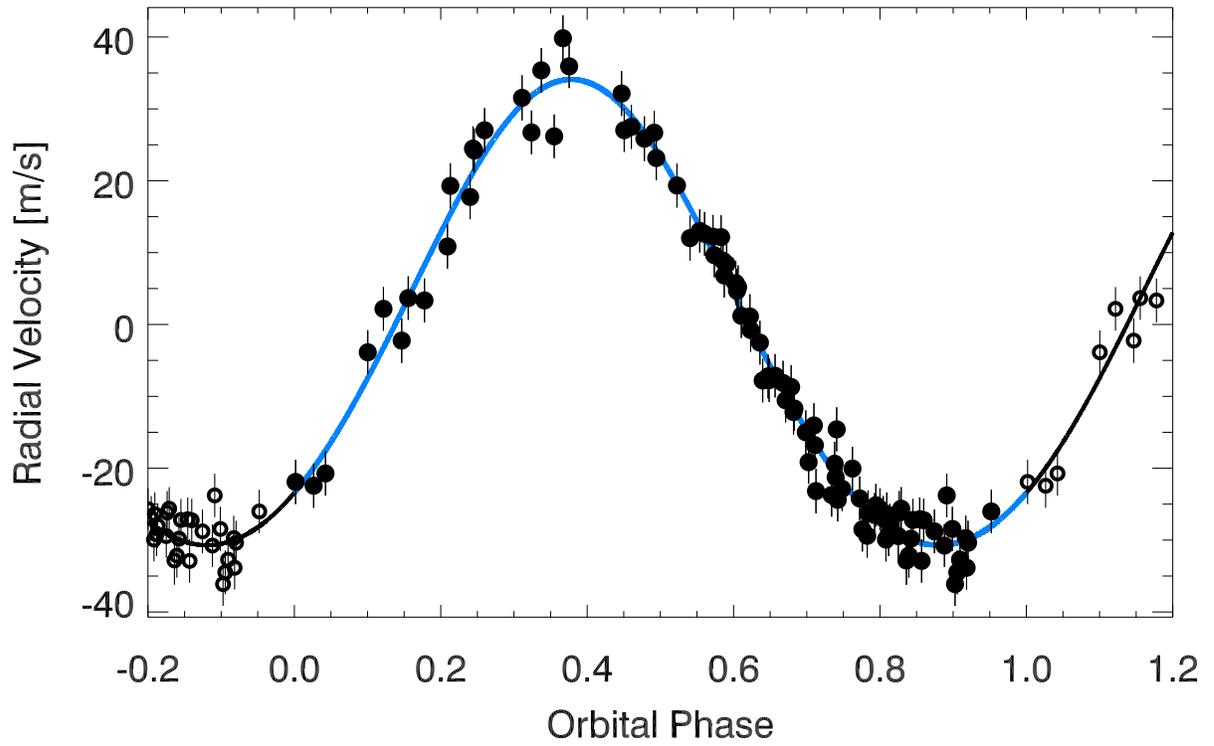}
\figcaption{The phase-folded radial velocities for HIP~57274c are shown with theoretical 
Keplerian velocities for the inner and outer planets and the linear trend subtracted. The Keplerian model 
has a period of $32.0 \pm 0.05$~d, eccentricity of $0.05 \pm 0.03$ and \msini = 130 \mearth\ or 0.41 \mjup.}
\end{figure}

\begin{figure}
\plotone{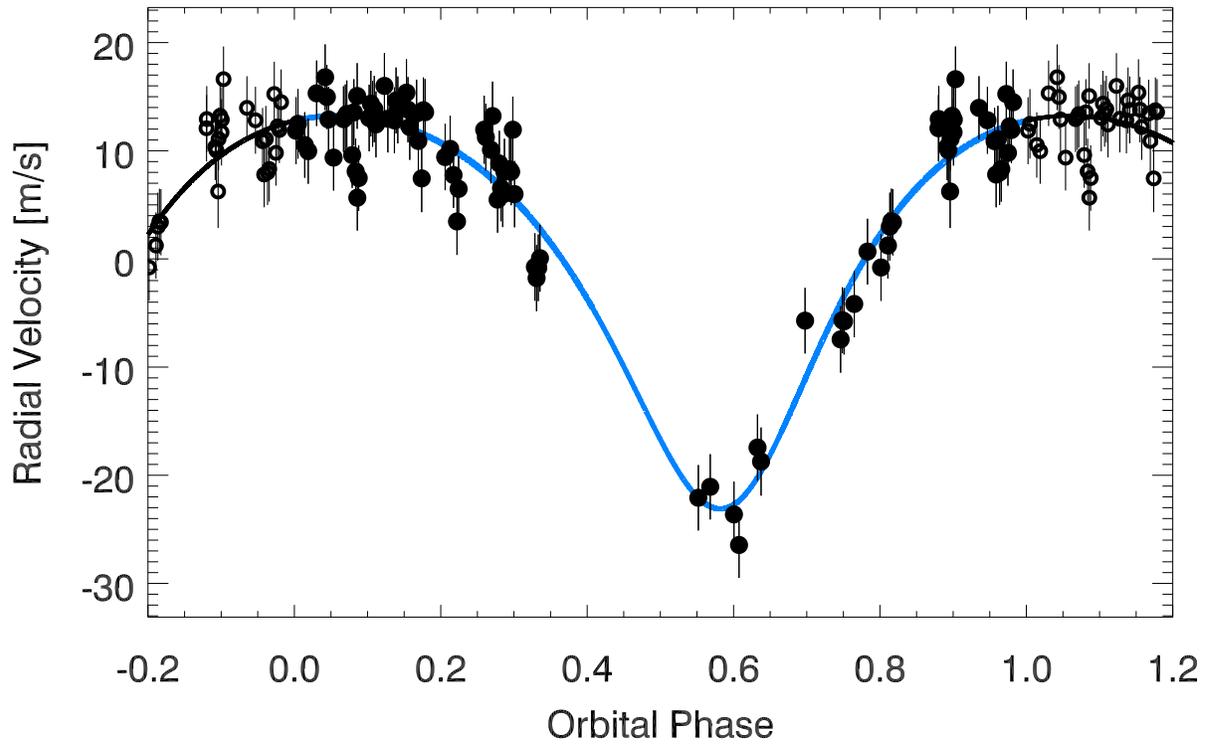}
\figcaption{The phase-folded data for HIP~57274d is shown with the two inner planets and a linear trend removed. 
The best fit Keplerian model has a period of $431.7 \pm 8.5$ d, \msini = 0.53 \mjup\ and an eccentricity of $0.27 \pm 0.05$.}
\end{figure}

\begin{figure} 
\plotone{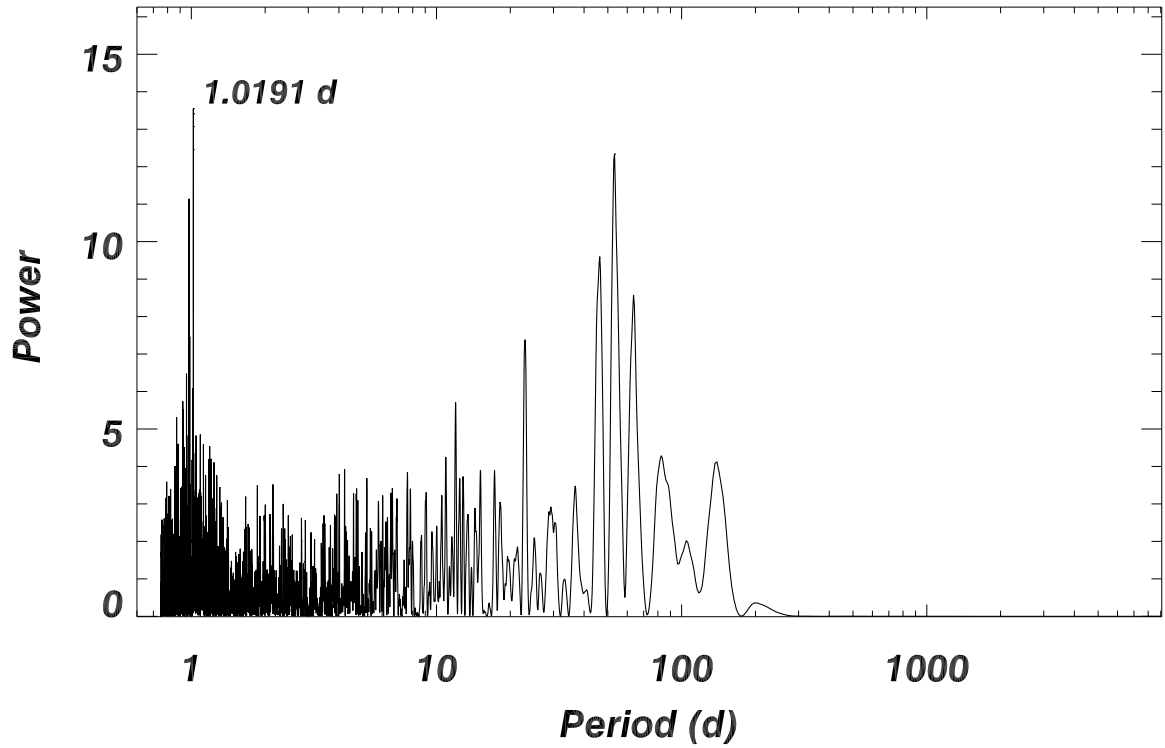}
\figcaption{After subtracting the linear trend and the best fit theoretical velocities for three planets, the periodogram 
of the residuals shows significant power near 1.0193 and 53 days.  These peaks are likely aliases of each other 
from the diurnal cadence, and the peak near 53 days could be associated with spots on the surface of this cool star. }
\end{figure}

\begin{figure}
\plotone{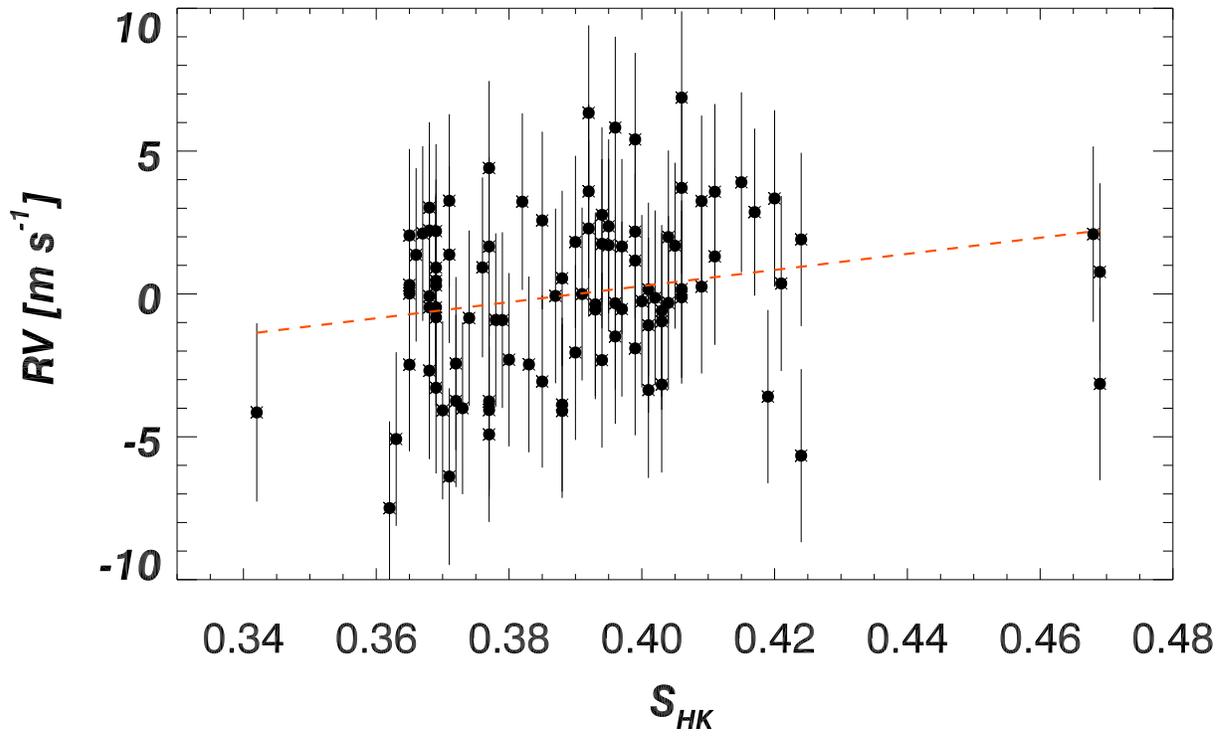}
\figcaption{The residual velocities to the triple planet model are plotted as a function of \shk\ activity measurements
and fit with a first order polynomial and do not show a significant trend.}
\end{figure}

\begin{figure}
\plotone{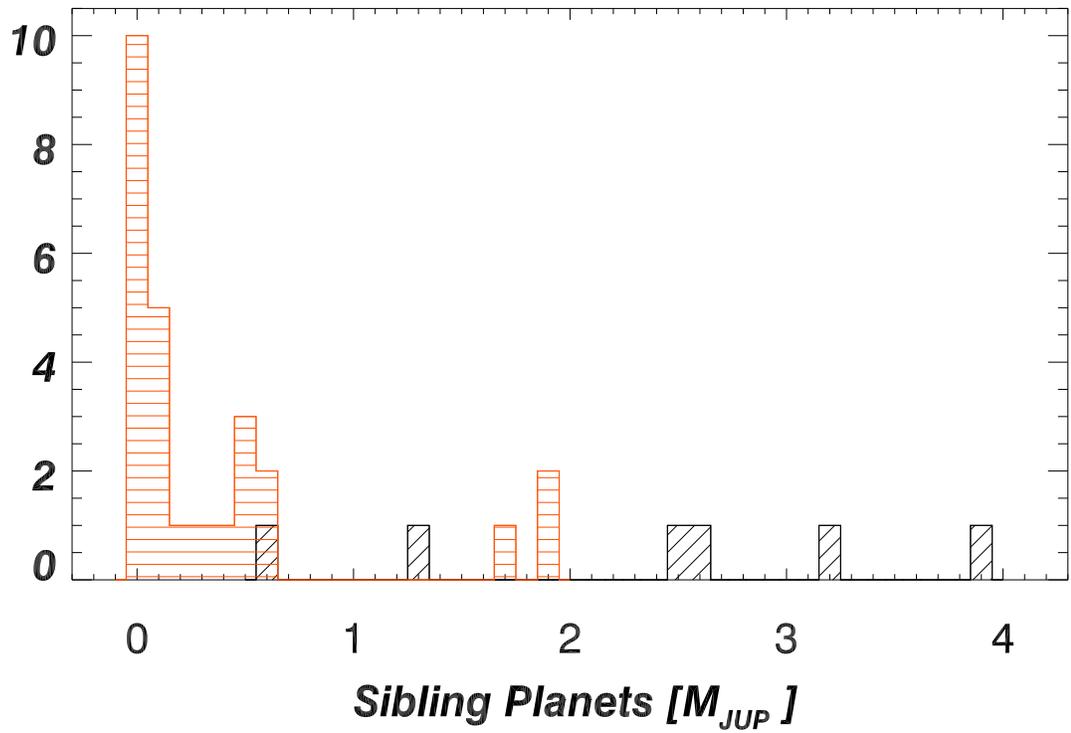}
\figcaption{Excluding the "primary" planets in our two subsamples, we show the distribution of sibling planets in the multi-planet 
systems with super-Earths  (red, horizontal hashed histogram) and jovian planets (black, diagonal hashed histogram).  }
\end{figure}

\begin{figure}
\plotone{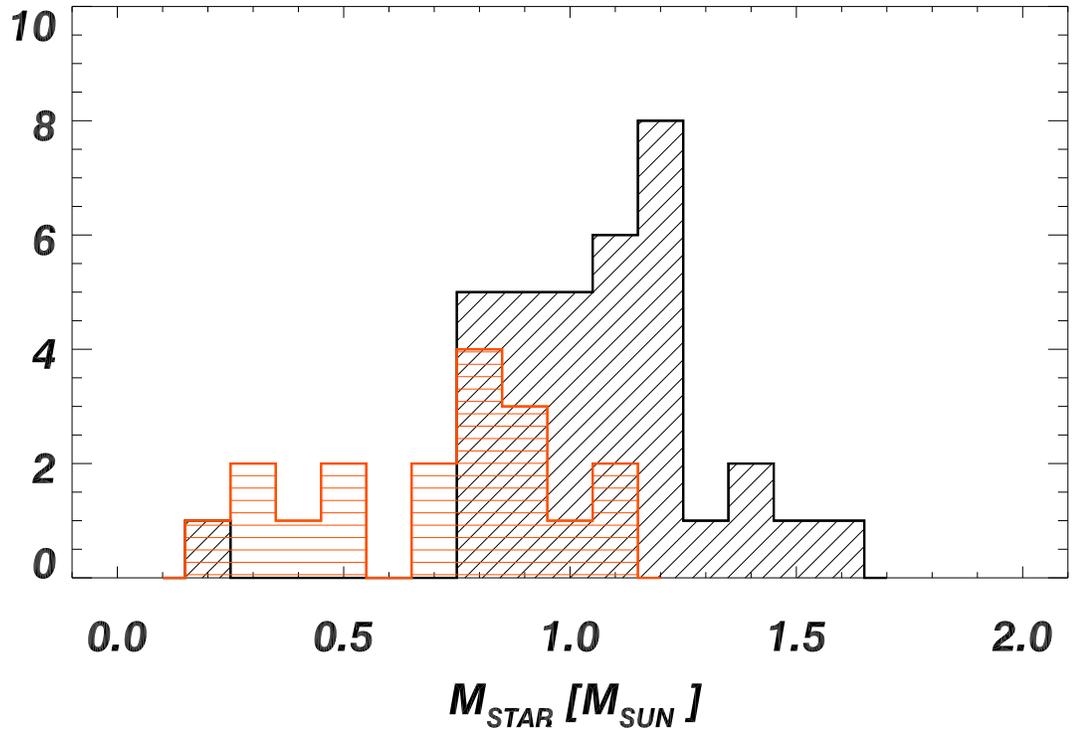}
\figcaption{Stellar hosts of detected super Earth planets (red, horizontal histogram) are systematically lower in 
mass than the stellar hosts of Jupiter mass planets (black diagonal lines). }
\end{figure}

\clearpage

\begin{deluxetable}{ll}
\tablenum{1}
\tablecaption{Stellar Parameters: HIP~57274}
\tablewidth{0pt}
\tablehead{\colhead{Parameter}  & \colhead{}  \\
}
\startdata
$V$                          & 8.96          \\
\bv                          & 1.11          \\
Spec Type                    & K5V           \\
Distance [pc]                & 25.92         \\
$M_V$                        & 6.89          \\
Bol Corr                     & -0.50         \\
$L_*$ [$L_\odot$]            & 0.19 (0.01)   \\
$M_*$ [$M_\odot$]            & 0.73 (0.05)   \\
$R_*$ [$R_\odot$]            & 0.68 (0.03)   \\
Age [Gyr]                    & 7.87 (5)      \\
\teff [K]                    & 4640 (100)    \\
\logg                        & 4.71 (0.1)    \\
\fe                          & +0.09 (0.05)  \\
\vsini [\kms]                & 0.5 (0.5)     \\
\shk                         & 0.39 (0.02)   \\
 \rhk\                       & -4.89         \\
$P_{rot}$                    & 45d           \\
Radial Velocity [\kms]       & -4.7          \\
\enddata                        
\end{deluxetable} 

\begin{deluxetable}{rrcc}
\tablenum{2}
\tablecaption{Radial Velocities for HIP~57274}
\tablewidth{0pt}
\tablehead{ 
    \colhead{}   &
    \colhead{RV} & 
    \colhead{$\sigma_{\rm RV}$} &
       \colhead{$S_{HK}$}  \\

       \colhead{JD-2440000} & 
       \colhead{(\ms)} & 
       \colhead{(\ms)} &
                                       \\
 }
\startdata
    14806.13297  &     0.65  &     1.34  &     0.469  \\
    14807.14718  &    -2.97  &     1.88  &     0.469  \\
    14809.15850  &    -1.80  &     1.26  &     0.468  \\
    15190.15800  &   -14.89  &     1.18  &     0.409  \\
    15198.16272  &    -7.39  &     1.29  &     0.403  \\
    15232.04776  &    11.50  &     1.26  &     0.411  \\
    15255.86086  &    -7.77  &     1.12  &     0.417  \\
    15255.96257  &    -7.31  &     1.28  &     0.415  \\
\enddata
\end{deluxetable}
\clearpage
 
\begin{deluxetable}{llll}
\tablenum{3}
\tablecaption{Orbital Parameters for HIP~57274}
\tablewidth{0pt}
\tablehead{\colhead{Parameter}  & \colhead{b}      & \colhead{c}         & \colhead{d}  \\
} 
\startdata
P (d)                             &  8.1352 (0.004)    & 32.03 (0.02)        & 431.7 (8.5)      \\
${\rm T}_{\rm p} - 2.44 x 10^6$ (JD)   &  14801.015 (1.3)   & 15785.208 (9.5)     & 15108.116 (14) \\
ecc                               &  0.187 (0.10)      & 0.05 (0.02)         & 0.27 (0.05)     \\
$\omega$ (deg)                    &  81 (59)           & 356.2 (120.0)       & 187.2 (5)        \\
K$_1$ (\ms)                       &  4.64 (0.47)       & 32.4 (0.6)          & 18.2 (0.5)       \\
$M\sin i$ (M$_{\oplus}$)          &  11.6 (1.3)        & 130 (3)             & 167.4  (8)       \\
$a_{rel}$ (AU)                    &  0.07              & 0.178               & 1.01                \\
$T_c$ (HJD - 2.44e6)              & 15801.779 (0.271)  & 15793.035 (0.176)   &                         \\
${\rm T_{duration}}$ (hr)         & 3.08 (0.35)        & 5.88 (0.1)          &                         \\
${\rm t_{prob}}$                  & 6.5\%              & 2.7\%               &                         \\
Trend (${\rm m\ s^{-1} d^{-1}}$)  &  -0.026 (0.002)    &                     &                         \\
Avg SNR                           &  225               &                     &                         \\
RMS (\ms)                         &  3.15              &                     &                          \\
${\rm Nobs}$                      &  99                &                     &                         \\
\rchisq\                          &  1.06              &                     &                         \\
Jitter [\ms]                      &  2.8               &                     &                         \\
\enddata                        
\end{deluxetable}                          
\clearpage


\end{document}